\documentclass[12pt]{article} 

\usepackage{arxiv}
\usepackage[utf8]{inputenc} 
\usepackage[T1]{fontenc}    
\usepackage{hyperref}       
\usepackage{url}            
\usepackage{booktabs}       
\usepackage{amsfonts}       
\usepackage{nicefrac}       
\usepackage{microtype}      
\usepackage{graphicx}       
\usepackage[authoryear,round]{natbib} 
\usepackage{doi}            
\usepackage{amsmath}        
\usepackage{bm}             
\usepackage{subcaption}     
\usepackage{multirow}       

\title{The Impact of Bitcoin ETF Approval on Bitcoin’s Hedging Properties Against Traditional Assets}

\author{
  Yihan Hong \\
  Olin Business School\\
  Washington University in St. Louis\\
  \And
  Hengxiang Feng \\
  Olin Business School\\
  Washington University in St. Louis\\
  \And
  Yinghan Wang \\
  Olin Business School\\
  Washington University in St. Louis\\
  \And
  Boxuan Li \\
  Olin Business School\\
  Washington University in St. Louis\\
}

\date{}

\renewcommand{\shorttitle}{\textit{arXiv} Bitcoin ETF Analysis}

\begin{document}
\maketitle

\begin{abstract}
The approval of the Bitcoin Spot ETF in January 2024 marked a transformative event in cryptocurrency markets, signaling increased institutional adoption and integration into traditional finance. This study examines Bitcoin’s changing relationships with traditional assets, including equities, gold, and fiat currencies, following this milestone. Using rolling correlation analysis, Chow tests, and DCC-GARCH models, we found that Bitcoin’s correlation with the S\&P 500 increased significantly post-ETF approval, indicating stronger alignment with equities. Its relationship with gold stabilized near zero, while its correlation with the U.S. Dollar Index remained consistently negative. These findings offer insights into Bitcoin’s evolving role in portfolios, implications for market stability, and future research opportunities on cryptocurrency integration into traditional financial systems.
\end{abstract}


\section{Introduction}

Bitcoin has historically been discussed as a potential hedge asset, theoretically similar to gold and government bonds, largely due to its decentralized nature and independence from traditional financial systems \cite{Bouri2017}. The premise that Bitcoin can serve as a safe haven is predicated on its historically low correlation with traditional assets such as stocks and bonds, particularly during periods of market turbulence. Several empirical studies have examined Bitcoin’s role in financial portfolios, concluding that it possesses the potential to hedge against inflation, currency devaluation, and economic crises \cite{Corbet2018}. However, unlike traditional physical assets, Bitcoin is digital and highly volatile, which has led to ongoing debates regarding its reliability as a consistent hedge \cite{Klein2018}. The central question remains whether Bitcoin can maintain this independence as it becomes increasingly financialized and integrated into global capital markets.

The regulatory environment for Bitcoin in the United States has undergone a significant evolution, culminating in a paradigm shift in early 2024. The U.S. Securities and Exchange Commission (SEC) initially approached cryptocurrencies with caution, focusing on prosecuting fraud in 2013 and rejecting the first Bitcoin ETF applications in 2017 due to concerns over market manipulation and lack of surveillance agreements with regulated markets. While a futures-based ETF was approved in 2021, the market long awaited a spot product that would allow for direct price exposure. The pivotal moment arrived in January 2024, when the SEC approved several Bitcoin Spot ETFs, allowing institutional investors to gain exposure to Bitcoin without direct ownership \cite{Gensler2024}. This approval introduced strict transparency and surveillance requirements, signaling Bitcoin’s maturation as a regulated financial asset \cite{Uyeda2024}.

This regulatory milestone provides a unique opportunity to re-evaluate Bitcoin's market behavior. As institutional capital flows into the market through regulated channels, the asset's correlation dynamics may fundamentally change. Existing literature suggests that integration into mainstream financial systems could erode Bitcoin's independence, potentially increasing its correlation with equities \cite{Wang2021}. This phenomenon is often observed in emerging asset classes that transition from niche to mainstream. This study aims to fill the research gap by systematically analyzing the impact of the ETF approval on Bitcoin’s correlation with traditional hedge assets (Gold, U.S. Dollar) and equities (S\&P 500). By employing a combination of rolling correlations, structural break tests, and dynamic conditional correlation models, we provide a comprehensive assessment of whether Bitcoin’s role as a "digital gold" or safe haven has been altered by its integration into traditional finance.

\section{Literature Review}

\subsection{Theoretical Foundation of Safe-Haven Assets}
The concept of safe-haven assets is rooted in behavioral finance theories from the 1990s. Tversky and Kahneman (1991) introduced reference-dependent models explaining that investors exhibit heightened sensitivity to losses during economic turbulence \cite{Tversky1991}. Building on this foundation, Baur and Lucey (2009) formally defined safe-haven assets as those that provide a hedging effect against systemic risks during extreme market volatility. These assets typically exhibit little to no correlation, or even a negative correlation, with the overall market trend during distress periods \cite{Baur2009}. Understanding the theoretical underpinnings of safe havens is crucial for evaluating whether a novel asset class like cryptocurrency can fulfill this role.

\subsection{Traditional Safe-Haven Assets and Limitations}
Scholars have traditionally identified gold, government bonds, and specific fiat currencies as mainstream safe havens. Baur and McDermott (2010) demonstrated that gold exhibits strong safe-haven properties in most developed markets during financial crises \cite{Baur2010}. Similarly, regarding government bonds, Flavin and Morley (2014) argued that long-term bonds meet the definition of safe-haven assets as they effectively reduce risk during turbulence, primarily due to the "flight to quality" phenomenon \cite{Flavin2014}. In the currency markets, Ranaldo and Söderlind (2010) identified the Swiss franc as a significant hedge against U.S. stock market downturns \cite{Ranaldo2010}. However, the effectiveness of these traditional assets is not static. Lucey and Li (2013) found that gold's properties vary over time, and recent studies by Chang and McAleer (2020) indicate that correlations between traditional safe havens and market volatility have increased in the post-COVID era, potentially diminishing their protective utility \cite{Chang2020}.

\subsection{Bitcoin: The "Digital Gold" Debate}
The academic community remains divided on Bitcoin's classification as a safe haven. Proponents, such as Bouri and Molnar (2017), refer to Bitcoin as "digital gold" due to its fixed supply and decentralization. Bouri and Shahzad (2020) further argued that Bitcoin offers superior independence compared to gold and commodities because it is not tied to any central bank or geopolitical entity \cite{BouriShahzad2020}. Conversely, skeptics highlight its extreme volatility as a disqualifying factor. Smales (2019) found that Bitcoin's lower liquidity and significantly higher volatility compared to traditional assets make it currently unsuitable as a safe haven \cite{Smales2019}. Furthermore, Choi and Shin (2022) observed that Bitcoin prices fluctuate significantly with financial uncertainty, behaving more like a speculative asset than a reliable hedge \cite{Choi2022}. This ongoing debate underscores the need to re-examine Bitcoin's properties specifically in light of recent market structural changes.

\subsection{The Impact of Bitcoin ETF Approval}
The approval of Bitcoin Spot ETFs represents a significant structural change that may resolve or exacerbate these debates. Gensler (2024) noted that this approval enhances the legitimacy of the Bitcoin market, potentially reducing the reputational risk for institutional investors. According to Foley \& Lardner LLP (2024), ETFs increase market transparency and security, simplifying the investment process for institutions \cite{Foley2024}. While this attracts liquidity, as noted by Uyeda (2024), it also integrates Bitcoin more tightly with traditional markets. Lee and Lee (2024) pointed out that while liquidity improved, the correlation between Bitcoin and traditional assets like the stock market increased post-approval \cite{Lee2024}. This suggests that the "financialization" of Bitcoin through ETFs might be altering its fundamental hedging characteristics, transitioning it from an idiosyncratic asset to a systemic one.

\section{Methodology}
\label{sec:methodology}

To thoroughly investigate the impact of the ETF approval, we adopt a multi-layered econometric approach. We begin with a visual inspection of trends, proceed to statistical verification of structural breaks, and conclude with a robust dynamic model to account for volatility clustering.

\subsection{Forward Rolling Correlation}
Our analysis begins with Forward Rolling Correlation to provide an intuitive, time-varying perspective on the relationship between Bitcoin and traditional assets. By calculating correlations over moving windows (30-day and 60-day), we can capture short-to-medium-term fluctuations that a static correlation measure would miss. The coefficient is calculated as:
\begin{equation}
\text{corr}_{xy,t} = \frac{\sum_{i=t-N+1}^{t} (x_i - \bar{x})(y_i - \bar{y})}{\sqrt{\sum_{i=t-N+1}^{t} (x_i - \bar{x})^2 \sum_{i=t-N+1}^{t} (y_i - \bar{y})^2}}
\end{equation}
where $N$ represents the window size. This method allows us to visually identify potential shifts in correlation trends around the event date, serving as a preliminary diagnostic tool for structural changes.

\subsection{Chow Test for Structural Breaks}
While rolling correlations offer visual evidence, they do not provide statistical confirmation of a regime change. To address this, we employ the Chow Test to scientifically detect structural mutations in the regression parameters before and after the ETF approval date (Jan 10, 2024). We apply this test to verify if the linear relationship ($Y_{BTC} = \alpha + \beta X + \epsilon$) underwent a statistically significant shift. The test statistic is defined as:
\begin{equation}
\text{Chow Statistic} = \frac{(RSS_{pooled} - (RSS_1 + RSS_2)) / k}{(RSS_1 + RSS_2) / (N_1 + N_2 - 2k)}
\end{equation}
A significant result here would confirm that the ETF approval acted as a structural breakpoint in the market dynamics, indicating that the fundamental pricing mechanism of Bitcoin relative to other assets has shifted.

\subsection{ARMA-DCC-GARCH Model}
Rolling correlations have limitations, such as the arbitrary choice of window size and the equal weighting of observations. Furthermore, financial time series, particularly cryptocurrencies, exhibit heteroskedasticity (volatility clustering). To overcome these limitations and provide a robust analysis of long-term dynamic correlations, we employ the Dynamic Conditional Correlation (DCC) GARCH model. First, we estimate univariate GARCH(1,1) models for each asset to standardize residuals. Then, we compute the dynamic correlation matrix $R_t$:
\begin{equation}
R_t = \text{diag}(Q_t)^{-1/2} Q_t \text{diag}(Q_t)^{-1/2}
\end{equation}
\begin{equation}
Q_t = (1 - a - b)\bar{Q} + a \epsilon_{t-1} \epsilon_{t-1}^T + b Q_{t-1}
\end{equation}
This model allows us to isolate the true correlation dynamics from the noise of volatility spikes, offering a clearer picture of how Bitcoin's integration with traditional markets has evolved post-ETF.

\section{Experiments}
\label{sec:experiments}

\subsection{Data Selection, Stationarity, and Descriptive Statistics}
This research analyzes daily adjusted closing prices for Bitcoin, the S\&P 500 (representing equities), Gold (a traditional safe haven), and the U.S. Dollar Index (DXY). The dataset, sourced from Yahoo Finance, is carefully segmented into two distinct periods centered around the ETF approval date of January 10, 2024. The **Pre-Event** period spans from October 1, 2023, to January 9, 2024, while the **Post-Event** period extends from January 11, 2024, to April 30, 2024. 

Before proceeding with time-series modeling, it is critical to ensure the data is stationary. We conducted the Augmented Dickey-Fuller (ADF) test on the excess returns of all four assets. As shown in Table \ref{tab:adf}, the ADF statistics for Bitcoin (-41.4453), S\&P 500 (-20.2299), Gold (-71.0724), and DXY (-71.6057) are all highly significant with p-values of 0.0000. This confirms that the return series are stationary, satisfying the prerequisite conditions for ARMA-GARCH modeling.

\begin{table}[ht]
    \centering
    \caption{ADF Test on Excess Return of Four Assets}
    \begin{tabular}{lcccc}
        \toprule
         & Bitcoin & S\&P 500 & Gold & DXY \\
        \midrule
        ADF Statistics & -41.4453 & -20.2299 & -71.0724 & -71.6057 \\
        p-value & 0.0000 & 0.0000 & 0.0000 & 0.0000 \\
        \bottomrule
    \end{tabular}
    \label{tab:adf}
\end{table}

The descriptive statistics, summarized in Table \ref{tab:descriptive}, reveal a dramatic shift in Bitcoin's market behavior following the regulatory approval. Post-event, Bitcoin's mean price surged from approximately 26,328 to 60,503, accompanied by a significant increase in volatility, with the standard deviation rising from 7,034.9 to 9,919.4. In terms of excess returns, Bitcoin showed marked improvement, shifting from a negative mean return of -0.0008 to a positive 0.0018. In stark contrast, traditional assets exhibited stability. The S\&P 500 saw a reduction in volatility, with its standard deviation decreasing from 280.8 to 182.9, while the U.S. Dollar Index remained the least volatile asset, reinforcing its role as a stable store of value.

\begin{table}[ht]
    \centering
    \caption{Descriptive Statistics of Closing Prices and Returns (Pre vs Post Event)}
    \resizebox{\textwidth}{!}{
    \begin{tabular}{lcccccccc}
        \toprule
        & \multicolumn{4}{c}{\textbf{Pre-Event (2023.10.01 - 2024.01.09)}} & \multicolumn{4}{c}{\textbf{Post-Event (2024.01.11 - 2024.04.30)}} \\
        \cmidrule(r){2-5} \cmidrule(l){6-9}
        Asset & Mean Price & Std Dev & Mean Ret & Std Ret & Mean Price & Std Dev & Mean Ret & Std Ret \\
        \midrule
        Bitcoin & 26,328 & 7,035 & -0.0008 & 0.0322 & 60,503 & 9,919 & 0.0018 & 0.0323 \\
        S\&P 500 & 4,183 & 280 & -0.0019 & 0.0171 & 5,141 & 182 & 0.0012 & 0.0070 \\
        Gold & 1,880 & 116 & -0.0020 & 0.0164 & 2,214 & 142 & 0.0012 & 0.0095 \\
        DXY & 104.9 & 3.1 & -0.0025 & 0.0132 & 104.5 & 0.9 & 0.0003 & 0.0032 \\
        \bottomrule
    \end{tabular}
    }
    \label{tab:descriptive}
\end{table}

\subsection{Rolling Correlation Analysis}
\begin{figure}[ht]
    \centering
    \begin{subfigure}{\textwidth}
        \centering
        \includegraphics[width=0.6\textwidth]{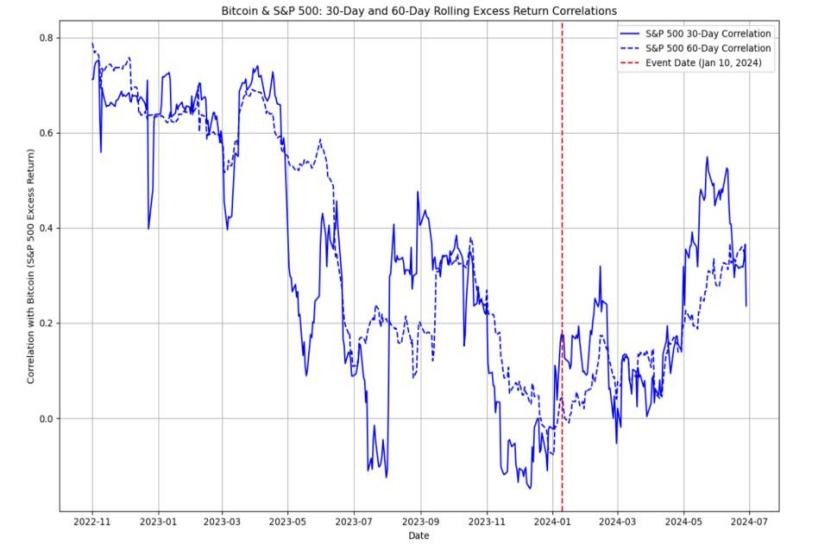}
        \caption{Bitcoin vs S\&P 500 (30-day \& 60-day)}
        \label{fig:roll_sp500}
    \end{subfigure}
    
    \vspace{1em}
    
    \begin{subfigure}{0.48\textwidth}
        \includegraphics[width=\textwidth]{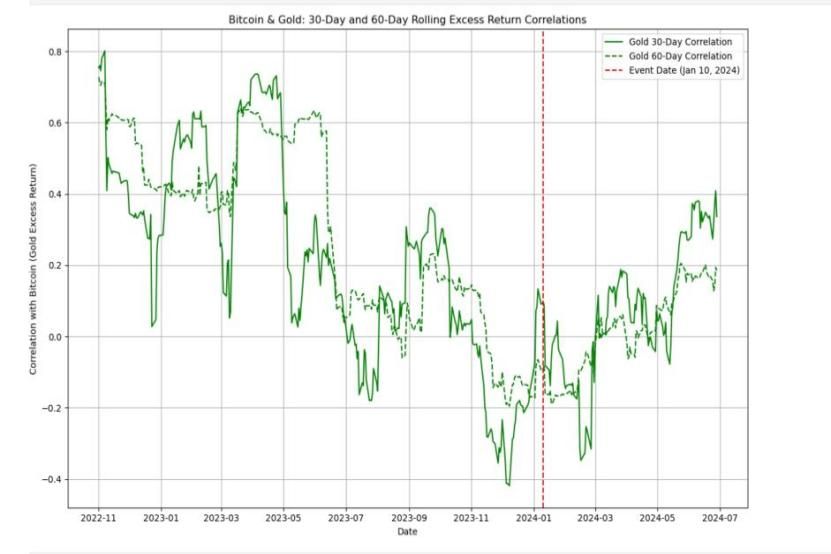}
        \caption{Bitcoin vs Gold}
        \label{fig:roll_gold}
    \end{subfigure}
    \hfill
    \begin{subfigure}{0.48\textwidth}
        \includegraphics[width=\textwidth]{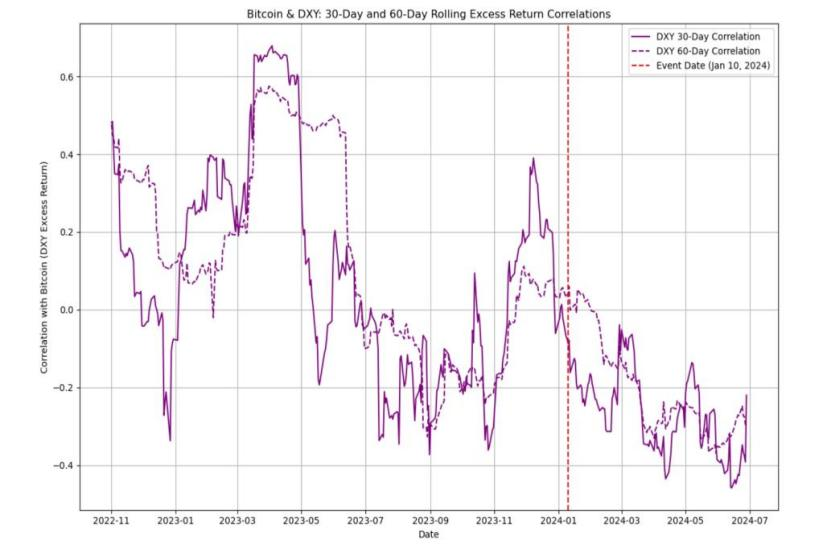}
        \caption{Bitcoin vs DXY}
        \label{fig:roll_dxy}
    \end{subfigure}
    \caption{Rolling Correlation Analysis across Assets.}
\end{figure}

To capture the dynamic evolution of market integration, we computed 30-day and 60-day rolling correlations. The results indicate a fundamental transition in Bitcoin's relationship with equities. Prior to the ETF approval, the correlation between Bitcoin and the S\&P 500 was volatile and generally trending downwards, often behaving idiosyncratically. However, immediately following January 2024, the correlation coefficient shifted to a sharp upward trend. This reversal suggests that the approval of the Bitcoin Spot ETF acted as a catalyst, transforming Bitcoin from an isolated asset into one that moves in tandem with traditional equities. As shown in Figure \ref{fig:roll_sp500}, the post-event trajectory implies that institutional investors are increasingly treating Bitcoin as a "risk-on" asset similar to technology stocks, rather than a diversified hedge.

Conversely, the relationship between Bitcoin and traditional safe havens did not exhibit a similar structural strengthening. The correlation with Gold (Figure \ref{fig:roll_gold}), while fluctuating, remained relatively low and did not show a significant positive shift after the ETF launch. This finding challenges the "digital gold" narrative, suggesting that despite its scarcity features, Bitcoin does not yet behave like a substitute for gold in institutional portfolios. Similarly, the correlation with the U.S. Dollar Index (Figure \ref{fig:roll_dxy}) remained consistently negative or negligible throughout the observation period. This indicates that Bitcoin's price discovery remains largely independent of fiat currency strength, even after its integration into regulated financial markets.

\subsection{Structural Break Verification (Chow Test)}
Visual inspection of correlation plots provides intuitive evidence, but statistical rigor is required to confirm regime changes. We employed the Chow Test to detect structural breaks at the January 10, 2024 breakpoint. The results revealed a nuanced and critical distinction between the stability of return coefficients and the structural shift in correlation trends. 

First, we examined the **Pairwise OLS Regression** on excess returns (Table \ref{tab:chow_pairwise}). Interestingly, for the S\&P 500, the Chow statistic (0.1427) yielded a p-value of 0.8670, indicating no structural break in the *linear regression coefficient* itself. However, for Gold (p=0.0991) and DXY (p=0.0002), the null hypothesis was rejected, confirming a break. This suggests that while the instantaneous beta of Bitcoin to the S\&P 500 might not have shifted abruptly on day one, the relationship with safe-haven assets did.

\begin{table}[ht]
    \centering
    \caption{Chow Test Results Based on Pairwise OLS Regression}
    \begin{tabular}{lccc}
        \toprule
        Independent Variable & Chow Statistics & P-value & Break Confirmed? \\
        \midrule
        S\&P 500 Excess Return & 0.1427 & 0.8670 & No \\
        Gold Excess Return & 2.3225* & 0.0991 & Yes \\
        DXY Excess Return & 8.7886*** & 0.0002 & Yes \\
        \bottomrule
    \end{tabular}
    \label{tab:chow_pairwise}
\end{table}

However, the most definitive evidence of a paradigm shift comes from the Chow Test on the **Rolling Correlation Series** (Table \ref{tab:chow_rolling}). Here, the results were unequivocal. The test statistics for Bitcoin's correlation with the S\&P 500 (139.42 for 30-day), Gold (135.96 for 30-day), and DXY (105.86 for 30-day) all yielded p-values of 0.0000, statistically significant at the 1\% level. This confirms that the *correlation structure*—representing the sustained, moving relationship between assets—underwent a massive mutation. While the pairwise regression captures a snapshot, the rolling correlation captures the trend, and this trend unmistakably broke from its historical pattern upon the ETF approval.

\begin{table}[ht]
    \centering
    \caption{Chow Test Results on Rolling Correlation Series (OLS with Constant)}
    \begin{tabular}{lccc}
        \toprule
        Dependent Variable & Chow Statistic & P-value & Structural Break? \\
        \midrule
        Corr BTC-S\&P 500 (30-day) & 139.42 & 0.0000 & Yes \\
        Corr BTC-S\&P 500 (60-day) & 269.30 & 0.0000 & Yes \\
        Corr BTC-Gold (30-day) & 135.96 & 0.0000 & Yes \\
        Corr BTC-DXY (30-day) & 105.86 & 0.0000 & Yes \\
        \bottomrule
    \end{tabular}
    \label{tab:chow_rolling}
\end{table}

\begin{figure}[ht]
    \centering
    \begin{subfigure}{0.32\textwidth}
        \includegraphics[width=\textwidth]{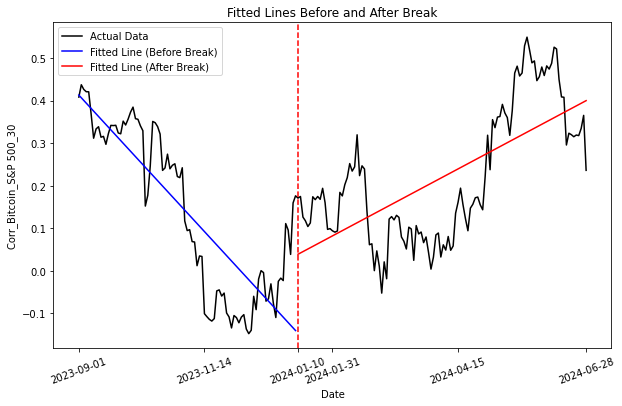}
        \caption{BTC vs S\&P 500}
        \label{fig:break_sp500}
    \end{subfigure}
    \hfill
    \begin{subfigure}{0.32\textwidth}
        \includegraphics[width=\textwidth]{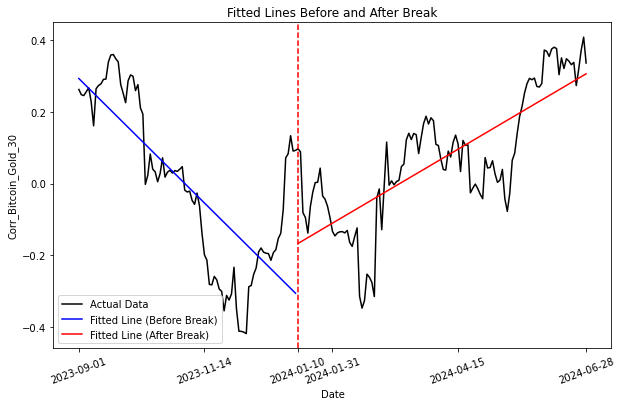}
        \caption{BTC vs Gold}
        \label{fig:break_gold}
    \end{subfigure}
    \hfill
    \begin{subfigure}{0.32\textwidth}
        \includegraphics[width=\textwidth]{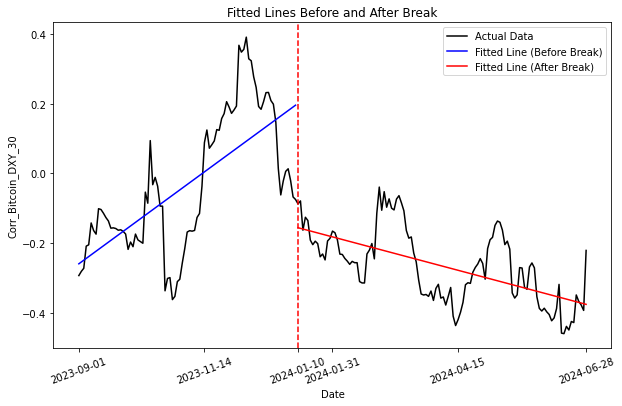}
        \caption{BTC vs DXY}
        \label{fig:break_dxy}
    \end{subfigure}
    
    \caption{Structural Break Visualizations (Chow Test). The red dashed line marks the ETF approval date. Note the upward trend shift for S\&P 500.}
    \label{fig:chow_breaks}
\end{figure}

The visualizations of these structural breaks (Figure \ref{fig:chow_breaks}) clearly show the divergence. The S\&P 500 correlation exhibits a distinct upward shift in the fitted line post-break, whereas the Gold and DXY correlations show different structural adjustments, further differentiating Bitcoin's integration with risk assets versus safe havens.

\subsection{Dynamic Conditional Correlation (DCC-GARCH) Analysis}
To account for the time-varying volatility inherent in cryptocurrency markets, we utilized the DCC-GARCH model. The estimation results for the ARMA-GARCH model (Table \ref{tab:garch_coef}) confirm the validity of this approach. The ARCH term ($\alpha_1 = 0.1184$) and GARCH term ($\beta_1 + \beta_2 \approx 0.49$) are significant, indicating the presence of volatility clustering which simple OLS would miss.

\begin{table}[ht]
    \centering
    \caption{Coefficients in the Volatility Equation of the ARMA-GARCH Model for Bitcoin}
    \begin{tabular}{lccc}
        \toprule
        Parameter & Coef & T Statistics & P-value \\
        \midrule
        Omega & 2.4025** & 2.197 & 0.0281 \\
        Alpha 1 & 0.1184* & 1.874 & 0.0609 \\
        Alpha 2 & 0.1683 & 1.454 & 0.146 \\
        Beta 1 & 0.1399 & 0.575 & 0.565 \\
        Beta 2 & 0.3563 & 1.588 & 0.112 \\
        \bottomrule
        \multicolumn{4}{l}{\footnotesize Note: ** denotes significance at 5\%; * denotes significance at 10\%.} \\
    \end{tabular}
    \label{tab:garch_coef}
\end{table}

The dynamic correlation results provide a sophisticated view of the post-ETF dynamics. Immediately following the approval, we observed a sharp drop in dynamic correlations across all asset pairs. This phenomenon can be attributed to the "dispersion effect", where the market experienced a short-term shock as investor structures adjusted and speculative capital rotated. The entry of new institutional participants initially caused divergent trading strategies, leading to this temporary decoupling as price discovery mechanisms reset.

\begin{figure}[ht]
    \centering
    \begin{subfigure}{0.48\textwidth}
        \includegraphics[width=\textwidth]{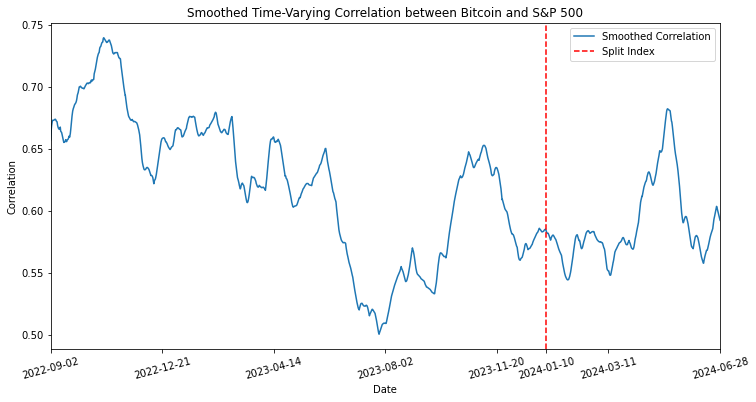}
        \caption{DCC: BTC vs S\&P 500}
        \label{fig:dcc_sp500}
    \end{subfigure}
    \hfill
    \begin{subfigure}{0.48\textwidth}
        \includegraphics[width=\textwidth]{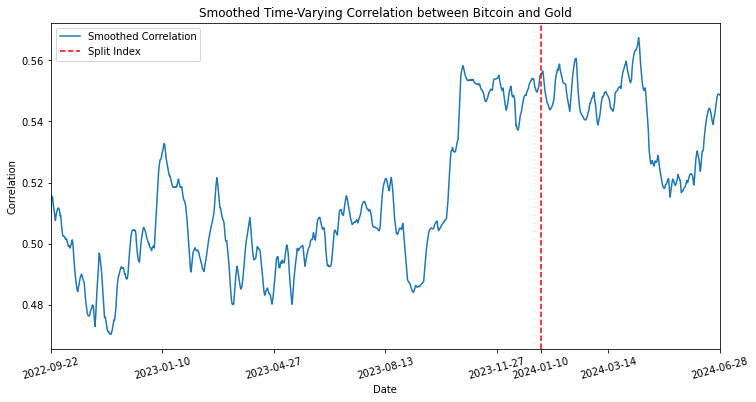}
        \caption{DCC: BTC vs Gold}
        \label{fig:dcc_gold}
    \end{subfigure}
    
    \vspace{1em}
    
    \begin{subfigure}{0.48\textwidth}
        \includegraphics[width=\textwidth]{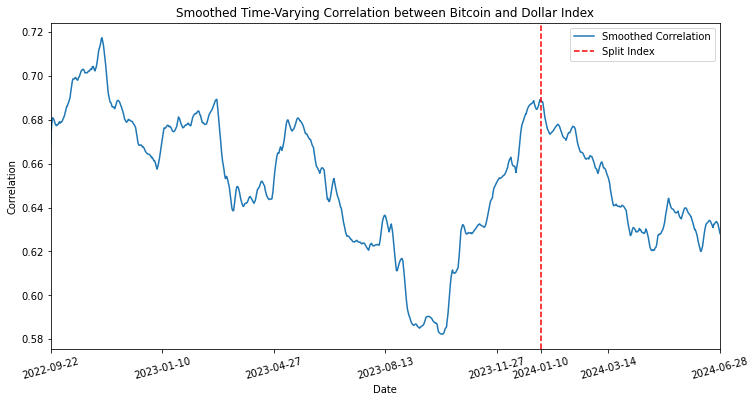}
        \caption{DCC: BTC vs DXY}
        \label{fig:dcc_dxy}
    \end{subfigure}
    
    \caption{Smoothed DCC-GARCH Dynamic Correlations showing post-ETF trends.}
    \label{fig:dcc_all}
\end{figure}

However, subsequent to this initial dispersion, the dynamic correlation between Bitcoin and the S\&P 500 rebounded significantly (Figure \ref{fig:dcc_sp500}). This rebound confirms the "portfolio reallocation" hypothesis, where investors, driven by optimistic sentiment, began to allocate capital simultaneously to both Bitcoin and equities. In contrast, the dynamic correlation with Gold (Figure \ref{fig:dcc_gold}) exhibited a mild downward trend. This supports the "fund diversion effect," where capital that might have previously sought refuge in gold is now being diverted to Bitcoin, but for speculative rather than hedging purposes. The competitive effect between these two assets suggests that Bitcoin's integration into the ETF framework is further decoupling it from gold's performance, weakening its potential utility as a safe haven in the traditional sense.

\section{Conclusion}

This study provides robust empirical evidence that the approval of the Bitcoin Spot ETF was a pivotal event that structurally altered Bitcoin’s role in financial markets. Our analysis confirms that the "financialization" of Bitcoin has led to a paradigm shift in its correlation dynamics. Specifically, the significant increase in Bitcoin's correlation with the S\&P 500 indicates a transition from an independent, speculative hedge to a conventional risk asset that moves in tandem with global equity markets. Conversely, the relationship with Gold and the U.S. Dollar remained stable or independent, challenging the popular narrative that the ETF approval would solidify Bitcoin's status as "digital gold". The statistical confirmation of structural breaks suggests that the pricing models for Bitcoin prior to 2024 may no longer be applicable in this new regulated regime.

For investors, these findings have profound implications for portfolio construction. The increasing correlation with equities suggests that Bitcoin's effectiveness as a diversifier against stock market downturns is diminishing. While it still offers high potential returns, it can no longer be relied upon as an uncorrelated safe haven during systemic equity shocks. Portfolio managers must now account for the higher beta and synchronization with the S\&P 500 when calculating Value at Risk (VaR) and optimal allocation weights. The data suggests that Bitcoin should be treated as a high-growth technology proxy rather than a defensive asset in the current market environment.

From a regulatory and systemic risk perspective, the integration of Bitcoin into mainstream finance introduces new channels for risk transmission. The stronger linkage between crypto markets and traditional finance means that volatility in one sector is more likely to spill over into the other. Policymakers should be aware that the "walled garden" of crypto is effectively gone; shocks in the crypto market now have a direct transmission mechanism to institutional balance sheets via ETFs. Future research should focus on the long-term persistence of these correlations and explore whether the "fund diversion effect" from gold becomes a permanent feature of the investment landscape.

\bibliographystyle{plainnat}

\end{document}